\documentclass{aastex61}

\received{November 23, 2016}
\revised{December 24, 2016}
\accepted{January 5, 2017}
\submitjournal{ApJL}

\shorttitle{Photogravitational assists at $\alpha$\,Centauri}
\shortauthors{Heller \& Hippke}

\begin{document}

\title{Deceleration of high-velocity interstellar photon sails into bound orbits at $\alpha$\,Centauri}

\correspondingauthor{Ren\'e Heller}
\email{heller@mps.mpg.de}

\author[0000-0002-9831-0984]{Ren\'e Heller}
\affil{Max Planck Institute for Solar System Research \\
\ Justus-von-Liebig-Weg 3 \\
\ 37077 G\"ottingen, Germany \\
\ \href{mailto:heller@mps.mpg.de}{heller@mps.mpg.de}
\textcolor{white}{.}\\
\textcolor{white}{.}\\}

\author{Michael Hippke}
\affiliation{Luiter Stra{\ss}e 21b \\
\ 47506 Neukirchen-Vluyn, Germany \\
\ \href{mailto:hippke@ifda.eu}{hippke@ifda.eu}}



\begin{abstract}

At a distance of about 4.22\,ly, it would take about 100,000 years for humans to visit our closest stellar neighbor Proxima Centauri using modern chemical thrusters. New technologies are now being developed that involve high-power lasers firing at 1\,gram solar sails in near-Earth orbits, accelerating them to 20\,\% the speed of light ($c$) within minutes. Although such an interstellar probe could reach Proxima 20 years after launch, without propellant to slow it down it would traverse the system within hours. Here we demonstrate how the stellar photon pressures of the stellar triple $\alpha$\,Cen\,A, B, and C (Proxima) can be used together with gravity assists to decelerate incoming solar sails from Earth. The maximum injection speed at $\alpha$\,Cen\,A to park a sail with a mass-to-surface ratio ($\sigma$) similar to graphene ($7.6\times10^{-4}\,{\rm gram\,m}^{-2}$) in orbit around Proxima is about $13,800\,{\rm km\,s}^{-1}$ ($4.6\,\%\,c$), implying travel times from Earth to $\alpha$\,Cen\,A and B of about 95 years and another 46 years (with a residual velocity of $1280\,{\rm km\,s}^{-1}$) to Proxima. The size of such a low-$\sigma$ sail required to carry a payload of 10 grams is about $10^5\,{\rm m}^2=(316\,{\rm m})^2$. Such a sail could use solar photons instead of an expensive laser system to gain interstellar velocities at departure. Photogravitational assists allow visits of three stellar systems and an Earth-sized potentially habitable planet in one shot, promising extremely high scientific yields.

\end{abstract}




\section{Introduction}
\label{sec:intro}

The discovery of Proxima\,b, a roughly Earth-mass planet \citep{2016Natur.536..437A} in the stellar habitable zone \citep{1993Icar..101..108K,2016arXiv160806919B,2016arXiv160808620M,2016A&A...596A.111R,2016A&A...596A.112T} around our closest stellar neighbor, has recently excited humanity's vision of interstellar travel to Proxima Centauri ($\alpha$\,Cen\,C, or Proxima). The Voyager 2 space probe, launched in 1977 using chemical thrusters, required about 12 years to reach Neptune and about 35 years to reach the boundary of the solar system \citep{2013Sci...341..150S} at a speed of $17\,{\rm km\,s}^{-1}$ relative to the Sun. At that speed, this 800\,kg spacecraft would take about 75,000 years to reach Proxima\,b.

Advances in the design of light, extremely thin and highly reflective solar sails \citep{2015arXiv150609214S} and the development of ultracompact electronic devices could soon allow the construction of extremely light high-tech solar sails that could be sent to Proxima. While earlier studies of laser-pushed lightsails toward $\alpha$\,Cen involved sail masses of the order of tons and sail areas of the order of square kilometers \citep{1966Natur.211...22M,1967Natur.213..588R,1984JSpRo..21..187F}, ultrathin, highly reflective 1\,gram photon sails \citep{Macchi2010} are now being investigated for interstellar travel to $\alpha$\,Cen \citep{2016arXiv160401356L,2016arXiv160909506M}. With an intended interstellar speed of $0.2\,c$, however, they would traverse a distance equivalent to the Moon's orbit around the Earth in just about six seconds, with little time left for high-quality close-up exploration and posing huge demands to the imaging system.

To maximize the scientific yield of such a mission, we here explore the new possibility of using Proxima's companion stars $\alpha$\,Cen\,A and B as photogravitational swings to decelerate an incoming light sail and deflect it into a bound orbit around Proxima, and possibly Proxima\,b. Using two astrophysical effects, the photon pressures and the gravitational tugs of $\alpha$\,Cen\,A, B, and C, such a sail could be maneuvered through the system without the need for onboard fuel.

\section{Methods}

Our aim is to find the maximum possible injection speed ($v_{\infty,\rm max}$) of a sail coming from Earth to perform a sequence of at least one fly-by at $\alpha$\,Cen\,A, B, and C. We studied all permutations of stellar encounters and found that the optimal sequence starts with a photogravitational assist at $\alpha$\,Cen\,A at the time when stars A, B, and C are all in the same plane with the incoming sail. This configuration minimizes the deflection angle ($\delta$) required by the sail to reach the next star (B in this case) to $\delta\approx10^\circ$, and it maximizes the injection speed for the first encounter, thereby minimizing the travel time from Earth. Proxima is not located in the orbital plane of the AB binary, but for a distant observer all three stars align about every 79.91 years (the orbital period of the AB binary). From the perspective of an incoming probe from Earth, the alignment occurs near the time of the AB periastron, the next of which will take place on 2035 June 24 \citep{Beech2015}.

\subsection{Photogravitational Assists}

\begin{figure}[t]
\centering
\includegraphics[width=.33\linewidth]{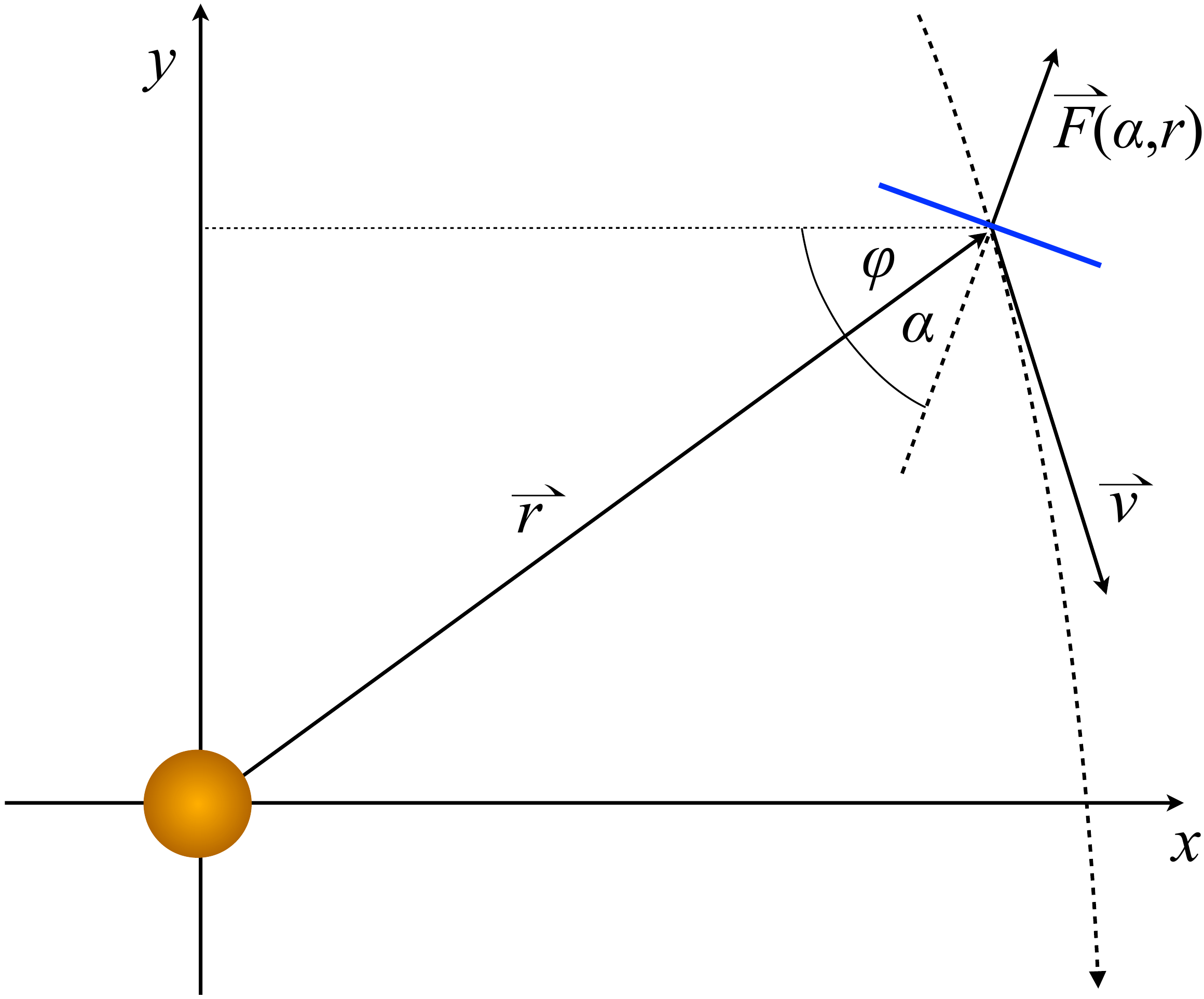}
\caption{Geometry of a slingshot trajectory. The star is in the origin of the coordinate system. The trajectory of the solar sail, with instantaneous velocity $\vec{v}$, is shown as a dotted curve in the $x$--$y$ plane. The pitch angle $\alpha$ between the normal to the sail plane and the radius vector to the star ($\vec{r}$) determines the stellar photon force $\vec{F}(\alpha,r)$ on the sail. \label{fig:geometry}}
\end{figure}

We consider a solar sail of mass $M$ and reflective surface area $A$ approaching a star at an instantaneous distance $\vec{r}$ and with an instantaneous velocity $\vec{v}$ in an $x$--$y$ plane (Figure~\ref{fig:geometry}). Its injection speed is $v_\infty$. The sail is subject to both the gravitational attractive force between it and the star and subject to the stellar radiation pressure, which, assuming a radially oriented sail and a uniformly bright, finite angular-sized stellar disc without limb darkening \citep{1990CeMDA..49..249M}, is

\begin{equation}
P(r) = \frac{\displaystyle L_\star}{\displaystyle 3{\pi}cR_\star^2} {\Bigg [} 1- {\Big [}1- {\Big (} \frac{\displaystyle R_\star}{\displaystyle r} {\Big )}^2{\Big ]}^{3/2} {\Bigg ]} \ \ ,
\end{equation}

\noindent
where $R_\star$ is the stellar radius and $L_\star$ is the stellar luminosity. We assume that the magnitude of the photon force on the sail is proportional to the photon pressure times the cosine of the pitch angle $\alpha$, hence

\begin{equation}\label{eq:Falpha}
F(\alpha,r) = \frac{\displaystyle L_\star A}{\displaystyle 3{\pi}cR_\star^2} {\Bigg [} 1- {\Big [}1- {\Big (} \frac{\displaystyle R_\star}{\displaystyle r} {\Big )}^2{\Big ]}^{3/2} {\Bigg ]} \times \cos(\alpha) \ \ .
\end{equation}

\noindent
We also explored the JPL numerical parametric force model \citep{1992spsa.book.....W}, where the reflectivity dependence on $\alpha$ is more complex, but these results agreed with those obtained via Equation~(\ref{eq:Falpha}) within 10 to 20\,\% of $v_{\infty,{\rm max}}$.

\subsection{Analytical Estimates of Maximum Deceleration}

Integration of Equation~(\ref{eq:Falpha}) with $\alpha~=~0^\circ$ over a sail path from $r_{\rm min}$ (the sail's minimum distance to the star) to $\infty$ yields the maximum possible reduction of kinetic energy from the photon pressure during a frontal approach:

\begin{equation}\label{eq:Ekin}
E_{\rm kin,p} =  {\displaystyle \int\limits_{r_{\rm min}}^{\infty} } dr F(r)_{|\alpha=0} = \frac{\displaystyle L_\star A}{\displaystyle 3{\pi}cR_\star^2} \ {\displaystyle \int\limits_{r_{\rm min}}^{\infty}} dr {\Bigg [} 1- {\Big [}1- {\Big (} \frac{\displaystyle R_\star}{\displaystyle r} {\Big )}^2{\Big ]}^{3/2} {\Bigg ]}\ \ .
\end{equation}

\noindent
For a full stop, Equation~(\ref{eq:Ekin}) is reduced by the kinetic energy gained by the sail through the conversion of potential into kinetic energy. Hence, the maximum possible reduction of the sail velocity for a full stop is given by

\begin{equation}\label{eq:vred}
v_{\rm red} =  \sqrt{\frac{\displaystyle 2 E_{\rm kin,p}}{\displaystyle M}}  - \sqrt{\frac{\displaystyle 2GM_\star}{\displaystyle r_{\min}}} \ \ .
\end{equation}

As an example, for an $M=1$\,gram sail with an area of $A=10\,{\rm m}^2$ that passes $\alpha$\,Cen\,A at $r_{\rm min}=5\,R_\star$ (assuming $R_\star=1.224$ solar radii and $L_\star=1.522\,L_\odot$; $L_\odot$ being the solar luminosity), we obtain $v_{\rm red}\,\approx\,1,200\,{\rm km\,s}^{-1}$, which corresponds almost precisely to the value of $v_{\infty,{\rm max}}$, achievable via a full stop trajectory as we found using our modified photogravitational trajectory code (see Section~\ref{sec:numerical}). The total maximum deceleration during a fly-by is slightly higher because the sail must climb out of the star's gravitational well after its passage, and therefore it slows down further.

\begin{figure*}[t]
\centering
\includegraphics[width=.485\linewidth]{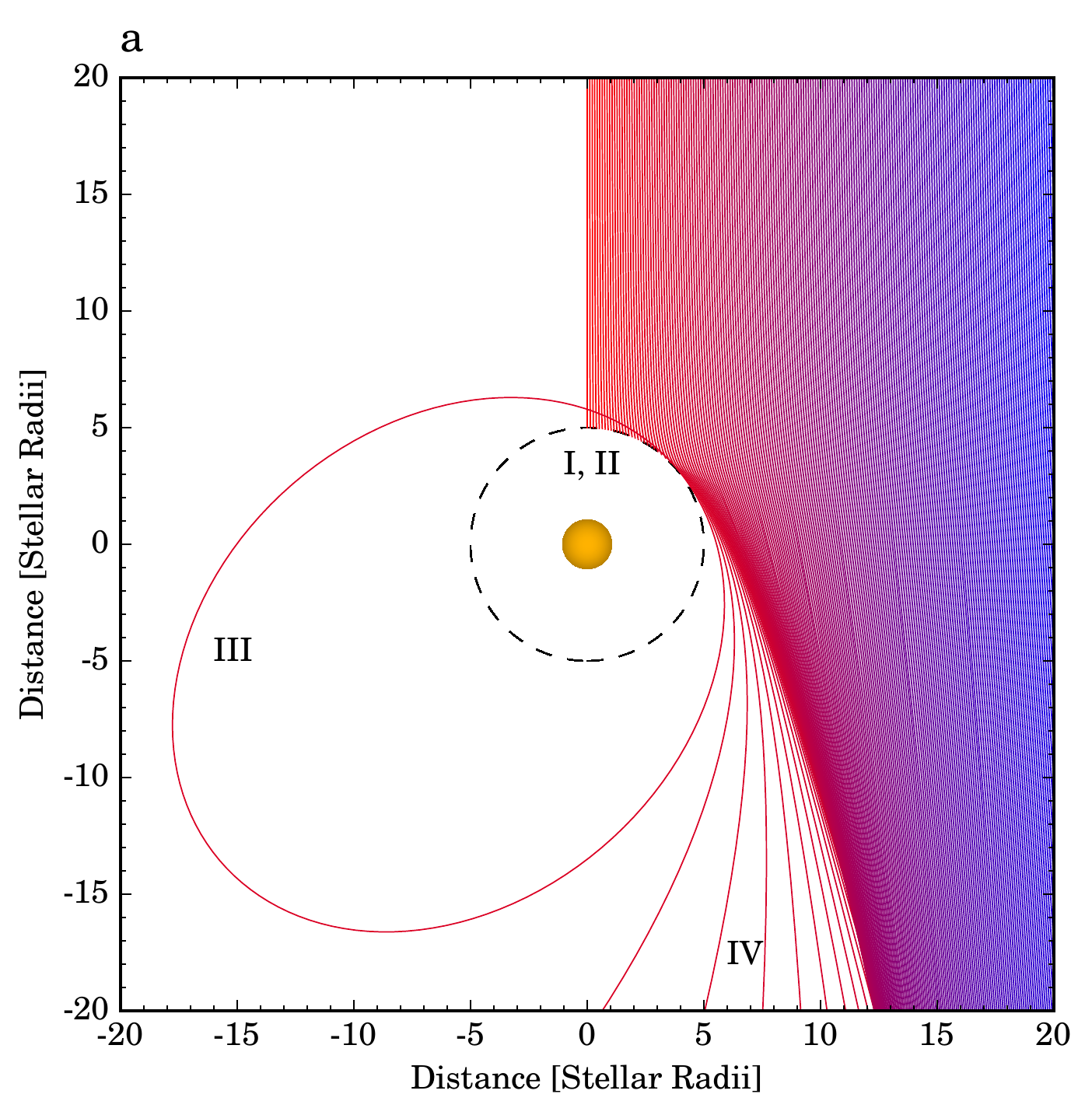}
\hspace{.2cm}
\includegraphics[width=.485\linewidth]{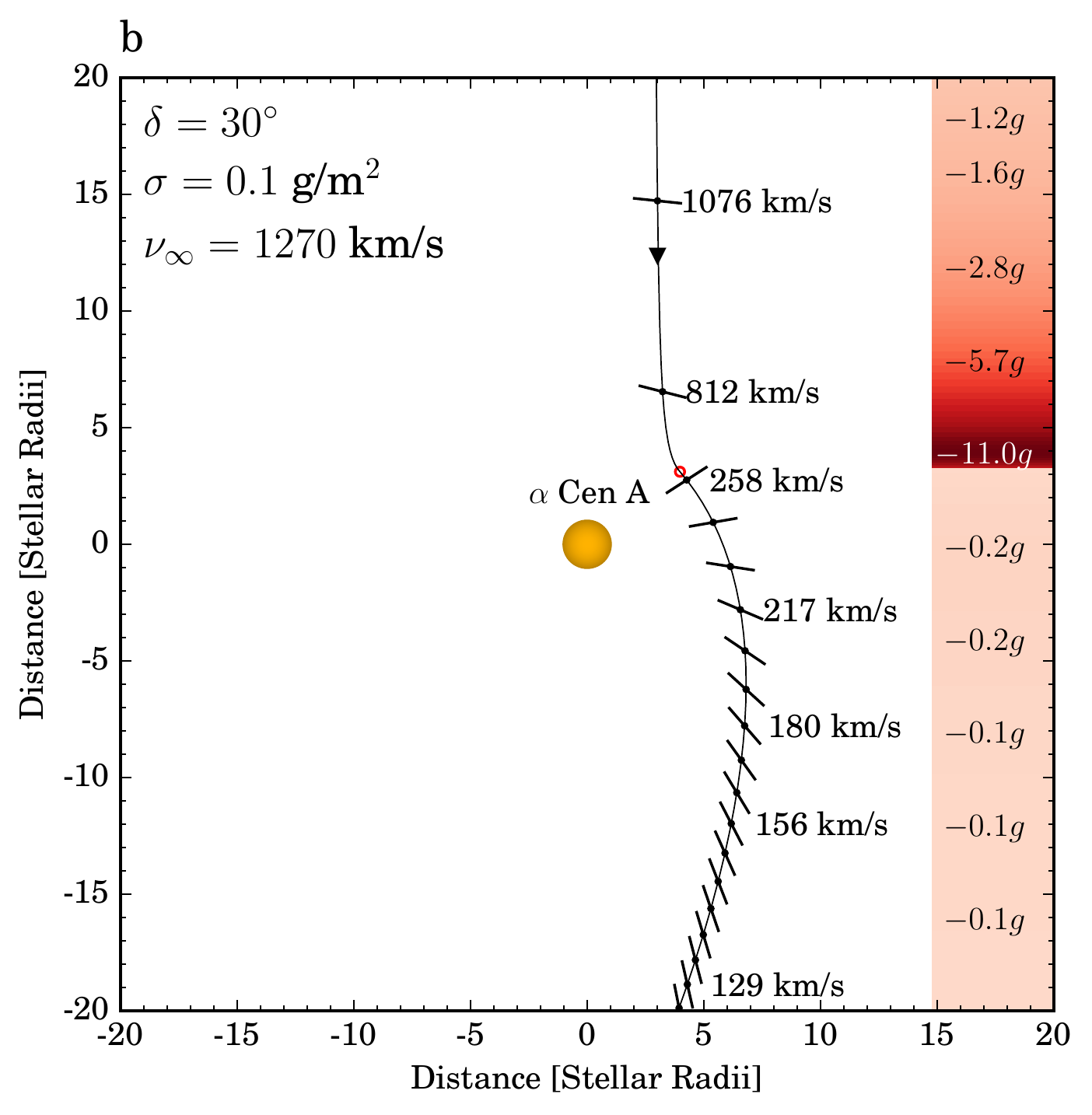}
\caption{Fly-by at $\alpha$\,Cen\,A. The orange star is located in the origin of the coordinate system. (a) The dashed circle indicates our nominal distance of closest approach, $r_{\rm min}=5\,R_\star$. The colored curves illustrate an iteration of our numerical computer code over various initial impact trajectories from Earth for a 1\,{\rm gram}/10\,m$^2$ sail with an initial speed of $1270\,{\rm km\,s^{-1}}$. Trajectory types I to IV are labeled. (b) Example trajectory of the same sail with its instantaneous speed shown along the trajectory in steps of 120 minutes. The sail's orientation is depicted by a black line, the direction of flight is indicated with an arrow at the top, and $r_{\rm min}$ is highlighted with a red open circle. The total time required by the sail to traverse this panel is 48\,hr. The sail's deflection angle, mass-per-area ratio, and injection speed for this example are shown in the top left corner. The instantaneous acceleration of the sail is shown in the color bar at the right, using units of the Earth's surface acceleration $g=9.81\,{\rm m\,s}^{-2}$. \label{fig:flyby} }
\end{figure*}

\begin{figure*}[t]
\centering
\includegraphics[height=0.478\textwidth]{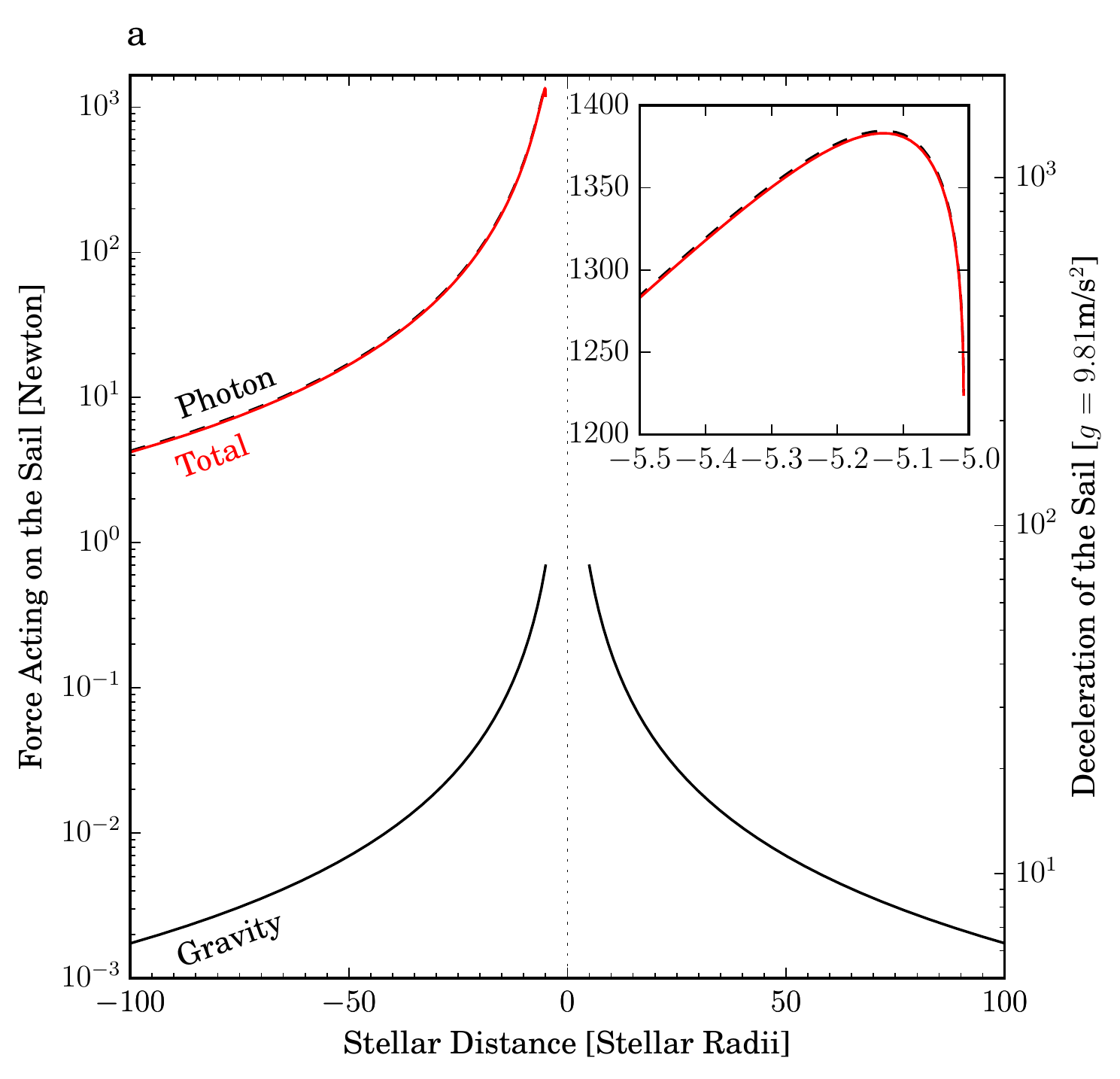}
\hspace{.3cm}
\includegraphics[height=0.486\textwidth]{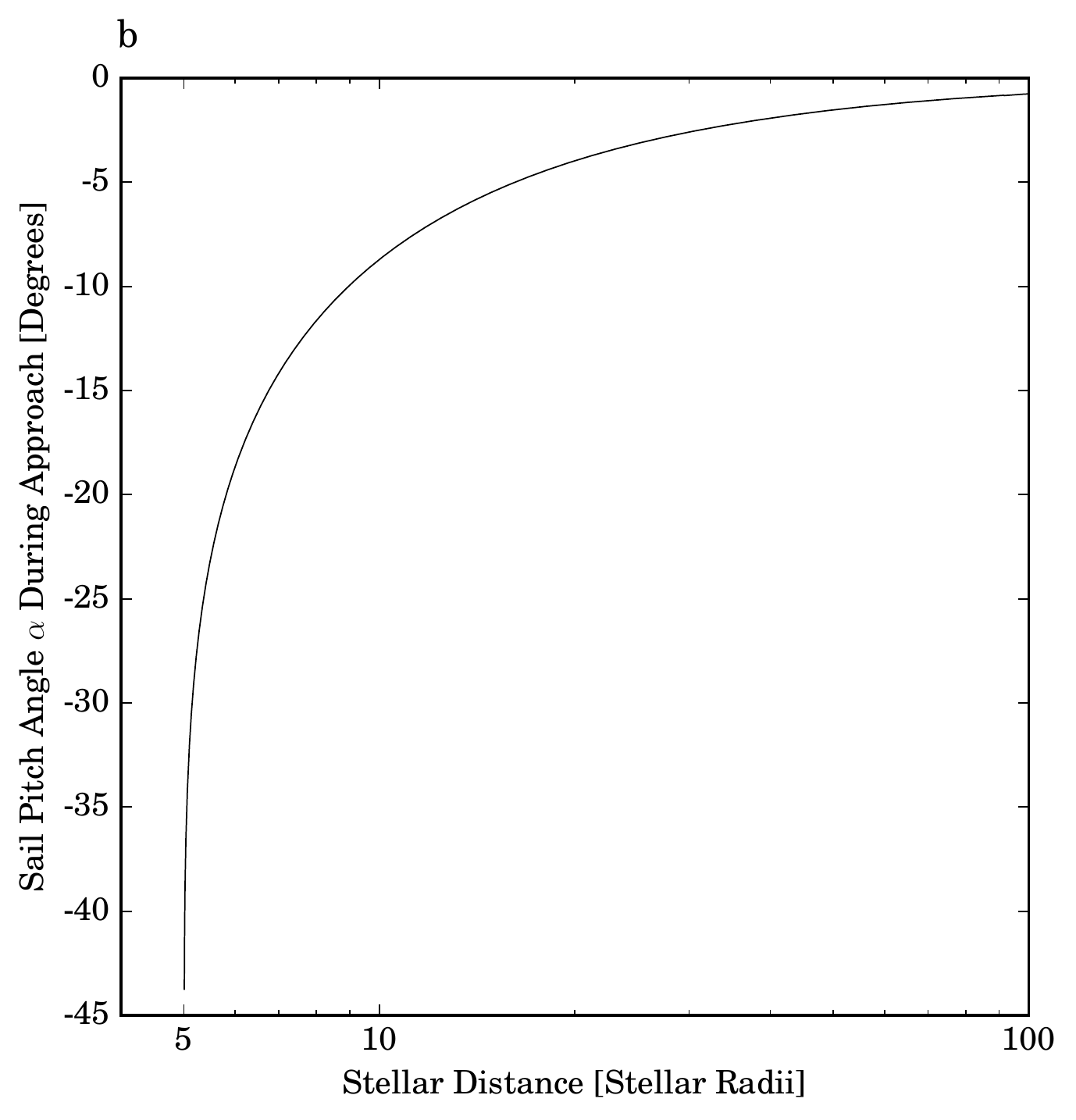}
\caption{(a) Forces (left ordinate) and deceleration (right ordinate) acting onto a $86\,{\rm gram}/(316\,{\rm m})^2=8.6\times10^{-4}\,{\rm gram\,m}^{-2}$ sail during a photogravitational assist with $v_{\infty}=v_{\infty,{\rm max}}=13,800\,{\rm km\,s}^{-1}$ at $\alpha$\,Cen\,A. Absolute values of the forces are shown for an approach as close as $r_{\rm min}=5\,R_\star$. The total force and deceleration during approach (i.e. photonic minus gravitational contributions) are indicated with a red solid line. During an approach from negative distances, photons decelerate the probe, whereas gravity is attractive and therefore accelerates it. After the encounter, the sail is rotated and aligned with the stellar radiation to avoid re-acceleration by the stellar photons. At this point, the total force becomes equal to the gravitational force and decelerates the sail. (b) Adjustment of the sail's pitch angle ($\alpha$) during the approach at $\alpha$\,Cen\,A. Solutions are derived numerically as per Equation~(\ref{eq:Fnu}) to maximize the sail's deceleration. \label{fig:approach} }
\end{figure*}

\subsection{Numerical Integrations of Sail Trajectories}
\label{sec:numerical}

We use a modified $N$-body code to model the gravitational pull by the stars under the additional effects of the stellar photon pressure onto the sail. As a key feature of our simulations, we solve the problem of continuously adjusting the sail's orientation to maximize the deceleration due to the photon pressure.\footnote{\citet{Cassenti1997} used a similar method to optimize the acceleration of possible solar sails leaving the solar system.} Such an adaptive orientation could be achieved if the sail were able to modify its reflectivity across its surface, e.g. using technologies akin to the nanocrystal-in-glass approach \citep{Llordes2013,Ma2016}. An asymmetric reflectivity distribution would induce a torque on the sail, which would start to rotate it \citep{Kislov2004,2016AdSpR..57.1147H}.

The ($x,y$) components of the photon force are calculated as per

\begin{equation}
\vec{F}(\alpha,r) = F(\alpha,r) {\Big (} \cos(\alpha+\varphi), \sin(\alpha+\varphi) {\Big )} \ \ ,
\end{equation}

\noindent
where $\varphi=\arctan(y/x)$ (see Figure~\ref{fig:geometry}). The maximum deceleration of the sail is achieved if the force component

\begin{equation}\label{eq:Fnu}
F_\nu = F(\alpha,r) \cos(\nu)
\end{equation}

\noindent
acting into the opposite direction of the sail's velocity vector is maximized, where

\begin{equation}\label{eq:nu}
\nu = \arccos {\Bigg (} \frac{\displaystyle F_x v_x + F_y v_y}{\sqrt{\displaystyle F_x^2 + F_y^2 } \sqrt{\displaystyle v_x^2 + v_y^2 }} {\Bigg )}
\end{equation}

\noindent
is the angle enclosed by $\vec{F}$ and $\vec{v}$. We enter Equation~(\ref{eq:nu}) into Equation~(\ref{eq:Fnu}) and numerically determine the value of $\alpha$ that yields maximum deceleration. The resulting trajectory is here referred to as a photogravitational assist, along which the photon pressure is used to reduce speed while both the photon pressure and the gravitational tug from the star determine the sail's deflection angle.

The orbits of the $\alpha$\,Cen stellar triple system are non-coplanar, so we simulate each fly-by in a new $x$--$y$ plane. For a broad range of $v_\infty$, we numerically integrate hundreds of sail trajectories, each of which has a slightly different horizontal offset (in steps of $0.1\,R_\star$) in the $x$--$y$ plane.

Figure~\ref{fig:flyby}(a) illustrates an example of such a trajectory iteration for an $M=1$\,gram, $A=10\,{\rm m}^2$ sail approaching $\alpha$\,Cen\,A with $v_\infty=1270\,{\rm km\,s}^{-1}$. All trajectories with an initial $x$-offset ${\lesssim}~3\,R_\star$ lead to a physical encounter of the sail with the star, i.e., the sail is lost (type I trajectory). For values of $v_\infty\lesssim1200\,{\rm km\,s}^{-1}$, the stellar photon pressure can stop the sail beyond $5\,R_\star$, a distance that we use as a fiducial value to prevent the sail from destruction (see Section~\ref{sec:discussion}). We refer to such a trajectory as a full stop (type II). A full stop route allows the sail to reorient itself at $r_{\rm min}$ to then leave the stellar system against gravity into arbitrary directions by means of the photon pressure. A type III trajectory leads the sail into a bound elliptical orbit around the star, from where it could then depart in an arbitrary direction. Type IV trajectories are what we refer to as photogravitational fly-bys. Figure~\ref{fig:flyby}(b) shows a detailed example of the optimum photogravitational assist around $\alpha$\,Cen\,A toward $\alpha$\,Cen\,B with $v_\infty=1270\,{\rm km\,s}^{-1}$, with the instantaneous speed of the sail shown along its trajectory and the instantaneous acceleration shown in the color bar at the right.

As an important feature of this trajectory, note that the photon pressure acts to initially accelerate the sail toward positive $x$ values, that is, in the direction opposite to the intended deflection. The additional effect due to the gravitational force, however, can ultimately deflect the sail into very different directions. Somewhat larger deflection angles via photogravitational assists could be achieved if the trajectory were not determined according to the maximization of the instantaneous deceleration as per Eq. (6). Although this would come at the price of less efficient deceleration, future simulations could show if such a trade-off could effectively increase the maximum possible injection speed.

\section{Results}
\label{sec:results}

\begin{figure*}[t]
\centering
\includegraphics[width=.635\linewidth]{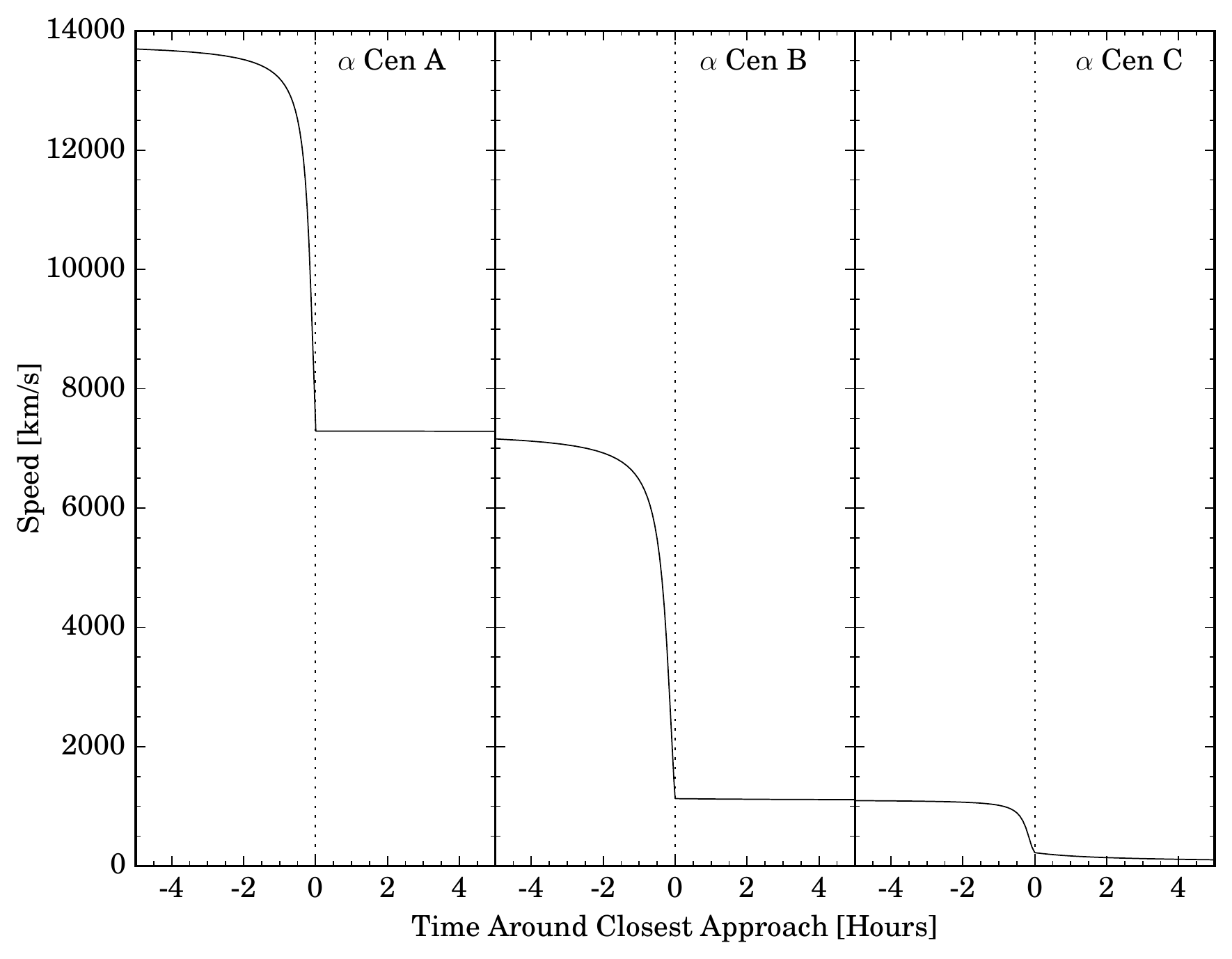}
\caption{Successive deceleration of our fiducial photon sail with a mass-to-surface ratio similar to graphene during photogravitational fly-bys at $\alpha$\,Cen\,A, B, and C. Its initial speed of $13,800\,{\rm km\,s}^{-1}$ is ultimately reduced to zero relative to Proxima, enabling stationary orbits around the star. \label{fig:ABC}}
\end{figure*}

The maximum injection speed at $\alpha$\,Cen\,A for a photogravitational assist to $\alpha$\,Cen\,B and Proxima is strongly dependent on the sail's mass-to-surface ratio ($\sigma$): the lower the $\sigma$ (for a given $M$) the stronger the photon force on the sail, and hence, the stronger the deceleration. We examined a range of plausible values of $\sigma$ and found that the lightest and most tensile materials known from laboratory experiments, such as graphene \citep[$\sigma=7.6\times10^{-4}\,{\rm gram\,m}^{-2}$;][]{Peigney2001507}, allow injection speeds of several percents of the speed of light. In order for such a sail to carry a payload (science, navigation, and communication instruments, reflective coating, etc.) of about 10 grams, the sail area must be on the order of $10^5\,{\rm m}^2$. We thus focus our analysis on a fiducial $\sigma=86\,{\rm gram}/(316\,{\rm m})^2=8.6\times10^{-4}\,{\rm gram\,m}^{-2}$ sail for which our numerical trajectory iteration yields $v_{\infty,{\rm max}}=13,800\,{\rm km\,s}^{-1}$.

One key issue for a high-velocity interstellar sail is in the extreme accelerations, which occur either during the departure from the solar system, or as in our case, during the arrival at the target star. Figure~\ref{fig:approach}(a) shows the contributions from the gravitational and photonic forces (left ordinate) or decelerations (right ordinate, in units of the Earth's surface acceleration $g~=~9.81\,{\rm m\,s}^{-1}$) on our $\sigma=8.6\times10^{-4}\,{\rm gram\,m}^{-2}$ sail during its photogravitational assist at $\alpha$\,Cen\,A. We find that the maximum forces are about $1400$\,N and the maximum accelerations are roughly $1600\,g$. Figure~\ref{fig:approach}(b) visualizes the continuous adjustment of $\alpha$ as a function of distance to $\alpha$\,Cen\,A.

Figure~\ref{fig:ABC} illustrates the reduction of an initial speed of $13,800\,{\rm km\,s}^{-1}$ for our fiducial sail into a parking orbit around Proxima after successive photogravitational fly-bys at $\alpha$\,Cen\,A and B. Each assist has been constrained to $r_{\rm min}>5\,R_\star$, where a sail with 99.999\,\% (99.99\,\%) reflectivity \citep{2016arXiv160401356L} absorbs about $24\,{\rm W\,m}^{-2}$ ($242\,{\rm W\,m}^{-2}$), corresponding to an effective temperature of about 144\,K (256\,K).

\begin{figure*}[t]
\begin{flushleft}
\hspace{.8cm} {\bf a} \hspace{8.8cm} {\bf b}
\end{flushleft}
\centering
\includegraphics[width=.485\linewidth]{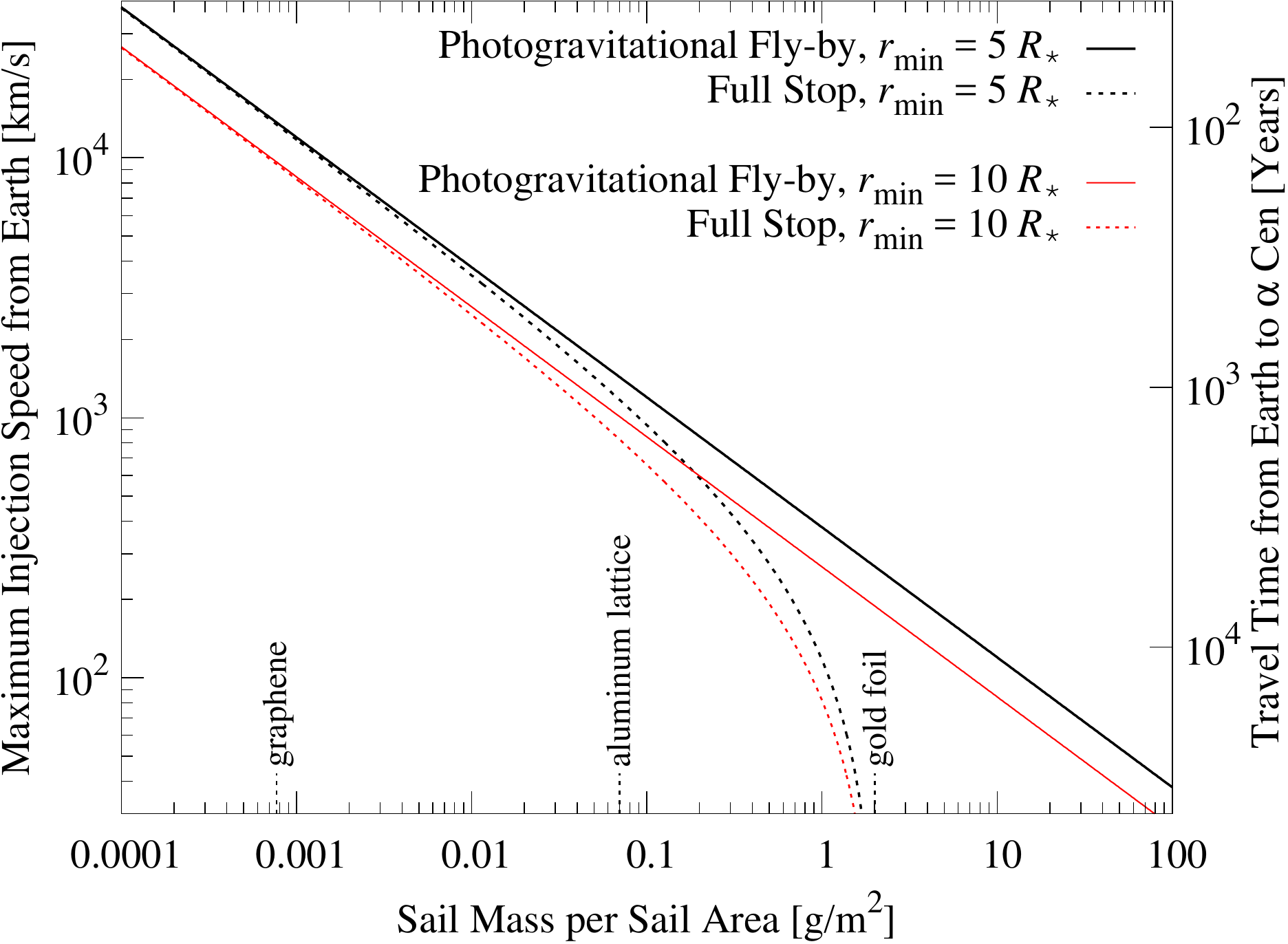}
\hspace{.2cm}
\includegraphics[width=.485\linewidth]{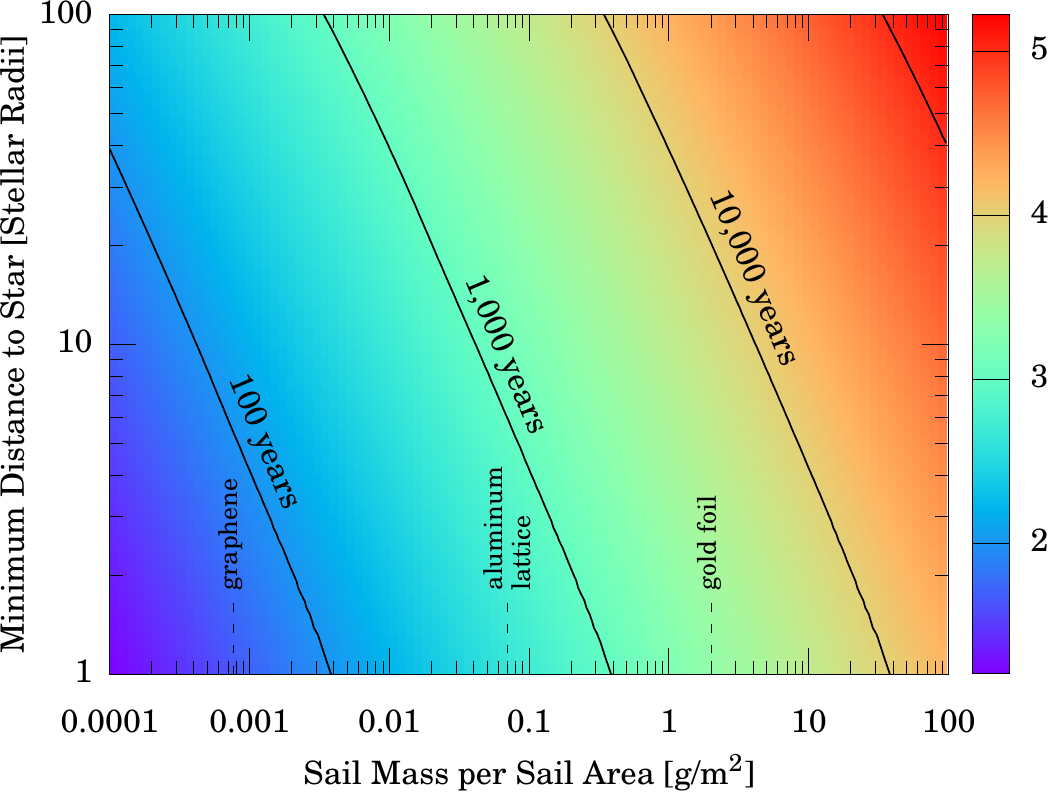}
\caption{Travel times of an interstellar sail to perform photogravitational assists at $\alpha$\,Cen\,A. All values derived as per Equation~(\ref{eq:vred}) and numerical integration of Equation~(\ref{eq:Ekin}) from $-\infty$ to $r_{\rm min}$. {\bf (a)} Maximum injection speeds ($v_{\infty,{\rm max}}$, left ordinate) and corresponding travel times (right ordinate) as a function of sail mass per sail area ($\sigma$). Black (red) lines refer to a minimum encounter distance with the star of $r_{\rm min}~=~5\,R_\star$ ($10\,R_\star$), with solid curves assuming a photogravitational fly-by and dashed curves assuming full stop trajectories. {\bf (b)} Contours of constant travel times for various values of $\sigma$ (abscissa) and $r_{\rm min}$ (ordinate). The color bar is in units of $\log_{10}({\rm time/years})$. \label{fig:times} }
\end{figure*}

One of the most critical aspects of a possible interstellar mission is flight duration. Figure~\ref{fig:times} shows the dependence of $v_{\infty,{\rm max}}$ (left ordinate) and the corresponding travel time (right ordinate) on $\sigma$ using analytical estimates for the maximum kinetic energy that can be absorbed using photogravitational fly-bys (solid lines) and full stops (dashed lines) near $\alpha$\,Cen\,A. Black lines assume $r_{\rm min}=5\,R_\star$ and red lines assume $r_{\rm min}=10\,R_\star$. We validated these values using our numerical $N$-body integrator and found that the analytical solution agrees with the numerical fly-bys within $<1\,\%$. Reference values of $\sigma$ for available materials or proposed ship designs are indicated at the bottom of the figure, suggesting that $v_{\infty,{\rm max}}=13,000\,{\rm km\,s}^{-1}$ (travel times of 100 years from Earth) could be achievable via photogravitational assists at $r_{\rm min}=5\,R_\star$ of our fiducial sail. After only a few days of travel between $\alpha$\,Cen\,A and B, such a sail could have maximum injection speeds of up to $1,280\,{\rm km\,s}^{-1}$ at Proxima to perform a transfer into a bound orbit via a full stop trajectory with $r_{\rm min}>5\,R_\star$. This means that the sail would take about 46 years to traverse the $\approx12,500$\,AU between the AB binary and Proxima \citep{2016arXiv161103495K}, which is very short compared to the travel time from Earth of about 1000 years at $1280\,{\rm km\,s}^{-1}$.

Figure~\ref{fig:times}(b) shows the travel times from Earth to $\alpha$\,Cen\,A derived from more than a million computations of $v_{\infty,{\rm max}}$ as per Equation~(\ref{eq:vred}) and using numerical integrations of Equation~(\ref{eq:Ekin}). An interstellar sail with a mass-to-surface ratio between graphene and an aluminum lattice ship \citep{Drexler1979} could arrive at $\alpha$\,Cen\,A within a few hundred years even if its closest approach to the star were restricted to $r_{\rm min}>10\,R_\star$.

\section{Discussion}
\label{sec:discussion}

\subsection{Limitations of the Model and Technical Challenges}

Close stellar encounters necessarily invoke the risk of impacts of high-energy particles and of thermal overheating. On the one hand, impacts of high-energy particles could damage the physical structure of the sail, its science instruments, its communication systems, or its navigational capacities. On the other hand, if those impacts could be effectively absorbed by the sail, they could even help to decelerate it. As shown in Section (\ref{sec:results}), heating from the stellar thermal radiation will not have a major effect on a highly reflective sail. However, the electron temperature of the solar corona is $>100,000$\,K at a distance of five solar radii. The \textit{Solar Probe Plus} (planned to launch in mid-2018) is expected to withstand these conditions for tens of hours \citep{2016SSRv..204....7F}, although the shielding technology for an interstellar sail would need to be entirely different \citep{2016arXiv160805284H}, possibly integrated into the highly reflective surface covering.

The effects of special relativity can be neglected in our calculations, with the Lorentz factor being \linebreak $1\leq\gamma=(1- v^2/c^2)^{-1/2}<1.005$ for $v<10\,\%\,c$. We also assumed that the stars are perfectly spherically symmetric. In reality, however, the acceleration of the probe will depend on the non-uniform mass distribution within the rotating star and possibly on effects from General Relativity \citep{2010AdSpR..46..346K}. Stellar oblateness and limb darkening affect the photon pressure by $<1\,\%$ at distances $>5\,R_\star$ \citep{1990CeMDA..49..249M}, and so these effects do not change the results of this study significantly. Yet, they must be considered for the planning of real missions.

Regarding the nautical issues of an A-B-C trajectory, communication among sails within a fleet could support their navigation during stellar approach, as it will be challenging for an individual sail to perform parallel observations of both the approaching star and its subsequent target star or of other background stars. Course corrections will need to be calculated live on board. In particular, the location of $r_{\rm min}$ will need to be determined on-the-fly because it will depend on the actual velocity and approach trajectory, and hence on the local stellar radiation pressure and magnetic fields \citep{2008A&A...489L..45R} along this trajectory.

\subsection{Further Applications of Photogravitational Assists}

Although we focused our simulations on the injection of light photon sails into bound orbits around Proxima, we also discovered that trajectory types II and III open up the opportunity for sample return missions to Earth (see Figure~\ref{fig:flyby}(a)). What is more, trajectory types II, III, and IV allow multi-fly-by missions at $\alpha$\,Cen\,A, B, and C. Among those, type IV will generally deliver $v_{\infty,{\rm max}}$.

Photogravitational assists can, of course, also be performed in the solar system. Once the technological implementation of a sail capable of photogravitational assists has been achieved, it seems natural to accelerate it to interstellar velocities using solar photons rather than using additional expensive technologies such as ground-based laser launch systems. Using Equation~(\ref{eq:vred}) and numerical integrations of Equation~(\ref{eq:Ekin}) from 5 solar radii to $\infty$, we estimate the maximum photonic ejection speed  from the solar system to be about $v_{\odot,{\rm max}}=11,500\,{\rm km\,s}^{-1}$ for an $8.6\times10^{-4}\,{\rm gram\,m}^{-2}$ sail, implying travel times to $\alpha$\,Cen of about 115 years. Beyond that, other nearby stars offer more favorable conditions than the $\alpha$\,Cen triple for the deceleration of incoming photon sails. Sirius\,A, as an example, at just about twice the distance from the Sun as $\alpha$\,Cen, offers a power of about $25\,L_\odot$ for deceleration. Consequently, the maximum possible injection speed of an $8.6\times10^{-4}\,{\rm gram\,m}^{-2}$ sail that can be absorbed ($44,600\,{\rm km\,s}^{-1}$ or $14.9\,\%\,c$) exceeds $v_{\odot,{\rm max}}$ by far. Thus, a laser launch system or alternative technologies would be required to accelerate photon sails from the solar system to these maximum possible speeds, maybe in combination with acceleration by the solar photon pressure.

In multi-stellar systems, successive fly-bys at the system members can leverage the additive nature of photogravitational assists. For multiple assists to work, however, the stars need to be aligned within a few tens of degrees along the incoming sail trajectory of the sail. Such a successive braking is particularly interesting for multi-stellar systems, where bright stars can be used as photon bumpers to decelerate the sail into an orbit around a low-luminosity star, such as Proxima ($0.0017\,L_\odot$) in the $\alpha$\,Cen system or the white dwarf Sirius\,B ($0.056\,L_\odot$) around Sirius\,A.

\section{Conclusions}

We present a new method of decelerating interstellar light sails from Earth at the $\alpha$\,Cen system using a combination of the stars' gravitational pulls and their photon pressures. This sailing technique, which we refer to as a photogravitational assist, allows multiple stellar fly-bys in the $\alpha$\,Cen stellar triple system and deceleration of a sail into a bound orbit. In principle, photogravitational assists could also allow sample return missions to Earth. The maximum injection speed to deflect an incoming, extremely light and tensile sail (with properties akin to graphene) carrying a payload of 10 grams into a bound orbit around Proxima is about $4.6\,\%\,c$, corresponding to travel times of 95 years from Earth. After initial fly-bys at $\alpha$\,Cen\,A and B, the sail could absorb another $1280\,{\rm km\,s}^{-1}$ upon the arrival at Proxima, implying an additional travel time between $\alpha$\,Cen\,AB and Proxima of 46 years.

Arrival at Proxima with maximum velocity could result in a highly elliptical orbit around the star, which could be circularized into a habitable zone orbit using the photon pressure near periastron. The time required for such an orbit transfer is small (years) compared to the total travel time. Once parked in orbit around Proxima, a sail could eventually use the stellar photon pressure to transfer into a planetary orbit around Proxima b.

In a more general context, photogravitational assists of a large, roughly $10^5\,{\rm m}^2=(316\,{\rm m})^2$-sized graphene sail could {\bf (1.)} decelerate a small probe into orbit around a nearby exoplanet and therefore substantially reduce the technical demands on the onboard imaging systems; {\bf (2.)} in principle allow sample return missions from distant stellar systems; {\bf (3.)} avoid the necessity of a large-scale Earth-based laser launch system by instead using the sun's radiation at the departure from the solar system; {\bf (4.)} limit accelerations to about $1000\,g$ compared to some $10,000\,g$ invoked for a 1\,m$^2$ laser-riding sail; and {\bf (5.)} leave of the order of 10 grams for the sail's reflective coating and equipment. These benefits come at the price of a yet to be developed large graphene sail, which needs to be assembled or unfold in near-Earth space and which needs to withstand the harsh radiation environment within $\gtrsim~5\,R_\star$ of the target star for several hours. This technical challenge, however, could be easier to tackle than the construction of a high-power ground-based laser system shooting laser sails in near-Earth orbits.

\acknowledgments
We thank the anonymous referee for a thoughtful report, and we thank Colin McInnes for his critical review of this manuscript prior to submission. This work has made use of NASA's Astrophysics Data System Bibliographic Services.

\software{\texttt{photograv} \citep{heller_rene_2017_236204}}. The software to create photogravitational trajectories is released\footnote{\url{www.github.com/hippke/photograv}, \texttt{commit \#727e77a}} under the free MIT license and can be used to reproduce the figures in this manuscript.




\end{document}